\documentclass[12pt]{iopart}

\usepackage{iopams}  
\usepackage{epsfig}

\begin{document}

\title{Early Time Evolution of High Energy Heavy Ion Collisions}

\author{Rainer J Fries}

\address{Cyclotron Institute and Department of Physics,
Texas A\&M University, College Station, TX 77843, USA}
\address{RIKEN/BNL Research Center, Brookhaven National Laboratory, Upton 
NY 11973}
\ead{rjfries@comp.tamu.edu}

\begin{abstract}
We solve the Yang-Mills equations in the framework of the McLerran-Venugopalan
model for small times $\tau$ after a collision of two nuclei. An analytic 
expansion around $\tau=0$ leads to explicit results for the field strength 
and the energy momentum tensor of the gluon field at early times. We then 
discuss constraints for the energy density, pressure and flow of the plasma 
phase that emerges after thermalization of the gluon field.
\end{abstract}

\submitto{\JPG}

The Color Glass Condensate (CGC) model 
\cite{MV:93,KMW:95,JMKMW:96,Kov:96,Gelis:07}
has provided valuable tools to understand the interaction of 
hadrons and nuclei at very high energies. The idea that the 
rapid growth of the gluon distribution at high energy is 
tamed by gluon recombination leads to a saturated gluon density
described by a saturation scale $Q_s$. Due to the high occupation numbers 
the gluons can be approximated by a classical field \cite{MV:93}. 
In high energy nuclear collisions the gluon density in each nucleus is 
enhanced by a factor $\sim A^{1/3}$ and it has been argued that the CGC 
can describe the initial particle production at the Relativistic 
Heavy Ion Collider (RHIC) \cite{KN:01}.

Here, we discuss what can be learned about the energy and momentum deposited
in the space between the two receding nuclei immediately after the collision.
This is an important question because there is convincing evidence that 
relativistic hydrodynamics governs the evolution of the 
fireball starting at rather early times  
$\tau_0 \approx 0.5 \ldots 1.0$ fm/$c$. 
The plasma phase, given by an energy momentum tensor 
$T^{\mu\nu}_{\mathrm{pl}} = (e+p)u^\mu u^\nu - p g^{\mu\nu}$ 
emerges through a rather rapid and not yet completely understood 
thermalization process from the initial gluon field after overlap, 
described by an energy momentum tensor $T^{\mu\nu}_{\mathrm{f}}$.
Our goal here is to constrain the energy density $e$, 
pressure $p$ and flow velocity $\mathbf{v}$ (with $u^\mu = 
\gamma(1,\mathbf{v})$) at the beginning of the plasma phase, using the 
CGC model in its version first conceived by McLerran and Venugopalan 
(MV) \cite{MV:93,KMW:95}.

We have to solve the Yang-Mills equations $[D_\mu, F^{\mu\nu}] = J^\nu$
for a current $J^\nu$ created by infinitely thin color charge distributions 
$\rho_1(\mathbf x_\perp)$ and $\rho_2(\mathbf x_\perp)$
propagating on the plus and minus light cone, respectively, and overlapping
at time $t=0$. In light cone gauge the field of each nucleus before the 
collision, $A^i_{1}(\mathbf x_\perp)$ and $A^i_{2}(\mathbf x_\perp)$ 
respectively, ($i=1,2$), is transverse. In the forward light cone,
i.e. after the collision, the field has light cone components
$A^\pm = \pm x^\pm A$ and transverse components $A^i$. $A$ and $A^i$
are functions of the proper time $\tau = \sqrt{t^2-z^2}$ and the transverse 
position $\mathbf{x}_\perp$, but not of the space-time rapidity $\eta$.

For small times $\tau$, immediately after the collision, the Yang-Mills 
equations in the forward light cone can be rewritten using an expansion
in powers of $\tau$, $A^\mu =\sum_n \tau^n A^\mu_{(n)}$. The resulting infinite
tower of differential equations can be solved recursively to arbitrary order 
in $\tau$ \cite{FKL:06}. 
Explicit solutions for $A^\mu$ and the field strength $F^{\mu\nu}$ up to 
order $\tau^3$ have been discussed. 
The lowest order in $\tau$ ($\mathcal{O}(\tau^0)$) coincides with the 
familiar boundary conditions for the field on the light cone. They 
imply that at the earliest time, for $\tau \to 0$, strong 
longitudinal electric and magnetic fields between the nuclei, 
\cite{FKL:05,FKL:06}
\begin{equation}
  E_z = ig [A_1^i,A_2^i], \qquad B_z = - ig \epsilon^{ij} [A_1^i,A_2^i],
\end{equation}
dominate. This was also observed, e.g., in \cite{Lappi:06,LMcL:06}. 
The next order in $\tau$ corresponds to a linear build-up of transverse 
fields.

One can now compute the energy momentum tensor of the gluon field.
The leading terms $\mathcal{O}(\tau^0)$ at small time are the diagonal
elements
\begin{equation}
  \varepsilon \equiv T^{00}_\mathrm{f} = T^{ii}_\mathrm{f} = 
  - T^{33}_\mathrm{f}
\end{equation}
($i=1,2$). Here $\varepsilon = (E_z^2 + B_z^2)/2$ denotes the energy
 density at $\tau = 0$. It turns out that $\varepsilon$ in the 
McLerran-Venugopalan model suffers from a ultraviolet divergence 
\cite{FKL:06,Lappi:06}. In lattice calculations this divergence is 
regularized by the finite lattice spacing \cite{Lappi:06}. 
We can argue that this divergence comes from the fact that the ultraviolet 
sector of the initial particle production with large momenta $p_\perp >> Q_s$
is not perfectly described in the MV model which works best for $p_\perp 
\approx Q_s$. Instead, we can choose a cutoff $Q_0$ and replace the 
classical field above $Q_0$ with a quantum but weak coupling description 
known to be well-behaved in this regime, i.e.\ perturbative QCD.

At order $\mathcal{O}(\tau^1)$ the components describing
transverse flow of energy receive their leading contributions
\begin{equation}
  T^{0i}_\mathrm{f} = \nu^i \cosh\eta, \qquad 
  T^{3i}_\mathrm{f} = \nu^i \sinh\eta,
\end{equation}
with the transverse flow vector
\begin{equation}
  \nu^i = -\frac{\tau}{4} \nabla^i \left( E_z^2 + B_z^2 \right)
\end{equation}
($i=1,2$). We will see below that this translates directly to the 
existence of transverse flow in the plasma phase immediately after 
thermalization.

Further corrections to the components of $T^{\mu\nu}_\mathrm{f}$ up to 
order $\mathcal{O}(\tau^3)$ can be easily computed. This gives a rather
accurate picture of the energy momentum tensor at times $\tau << 1/Q_s$. 
At $\tau \approx 1/Q_s$ the expansion might fail, however the asymptotic
behavior for $\tau \to \infty$ is known from a weak coupling expansion
\cite{KMW:95} and an interpolation can be used for intermediate values 
of $\tau$. This reproduces the time dependence of numerical solutions 
of the Yang-Mills equations, see e.g.\ \cite{Lappi:06}.

In any case, the classical field approximation is not expected to hold
for very large times anyway. Instead, the field is expected
to decay on a time scale $\sim 1/Q_s$, maybe through instabilities in
the color fields \cite{Strickland:07} or particle production in 
the background gluon field \cite{KT:05}. This is widely assumed, 
though not proved, to lead to a thermalized plasma within a rather short time. 

One can, however, use energy and momentum conservation to estimate
the energy momentum tensor of the plasma. Suppose the thermalization
is sufficiently rapid around a time $\tau_0$ so that the system can be 
approximately described by $T_\mathrm{f}^{\mu\nu}$ for 
$\tau \lesssim \tau_0$, and by $T_\mathrm{pl}^{\mu\nu}$ for 
$\tau \gtrsim \tau_0$. Then one can show that
\begin{eqnarray}
  e+p &= \epsilon \left( 1-\left(\frac{\nu}{\epsilon+p}\right)^2 \right), & \\
  v^i &= \frac{1}{\cosh\eta} \frac{\nu^i}{\epsilon+p}, \qquad\qquad\qquad 
  v^3 &= \tanh\eta 
\end{eqnarray}
($i=1,2$) using the first two terms in the $\tau$ expansion.
These four equations relate the five parameters $e$, $p$ and $\mathbf{v}$
in the plasma phase with components of the energy momentum tensor of the 
field. The system of equations can be closed by providing an additional
constraint, e.g. an equation of state. This result can be used as an initial
condition for a hydrodynamic evolution of the system.

Let us briefly discuss this result. First we notice that we recover 
boost-invariance for the longitudinal flow velocity 
$v^3$. Secondly, as mentioned before, there is initial radial flow $v^i$ 
in the plasma phase which is directly proportional to the energy flow $\nu^i$
of the field. For collisions with finite impact parameter this
also implies the existence of elliptic flow.

The results presented so far are functions of the single nucleus fields 
$A^i_{1,2}$ and have to be evaluated by putting the correct expressions for
those fields. They have been discussed in the literature, see e.g. 
\cite{JMKMW:96}. In \cite{FKL:06} estimates were presented in an abelianized
approximation in which the non-abelian effects were mimicked by
color screening at a typical distance $R_c = 1/Q_s$. To be precise, 
the screening radius $R_c$ at a given point in the transverse plane was
chosen to to be $R_c^{-2} = 4\alpha_s \sigma/3$ where $\sigma$ is the number
density of color charges in the nucleus \cite{FKL:06}. This approximation 
reproduces the gluon two-point function $\langle A^i(x) A^i(y)\rangle$ 
\cite{JMKMW:96} quite well. Thus we expect this approximation to give
a good estimate of the energy momentum tensor, since it has been shown
that $\varepsilon$ depends on $A^i_{1,2}$ solely through this two-point 
function \cite{Lappi:06}. Using this scheme one finds very simple formulae
for the center of two very large nuclei colliding head-on,
\begin{eqnarray}
  \varepsilon & = \frac{2\pi \alpha_s^3}{N_c} \sigma_1\sigma_2 \ln(1+c\zeta^2), 
   \\
  \nu^i &= -\tau\frac{ \pi \alpha_s^3}{2 N_c} \nabla^i (\sigma_1\sigma_2)
  \ln(1+c\zeta^2) ,
\end{eqnarray}
where $c\approx 0.42$ is a numerical constant and $\zeta = R_c Q_0$. 
$\sigma_{1,2}$ are the number densities of charges in
the two nuclei, which lead to the $SU(3)$ color densities 
$\rho_{1,2}$ \cite{FKL:06}.

\begin{figure}
\begin{center}
\epsfig{file=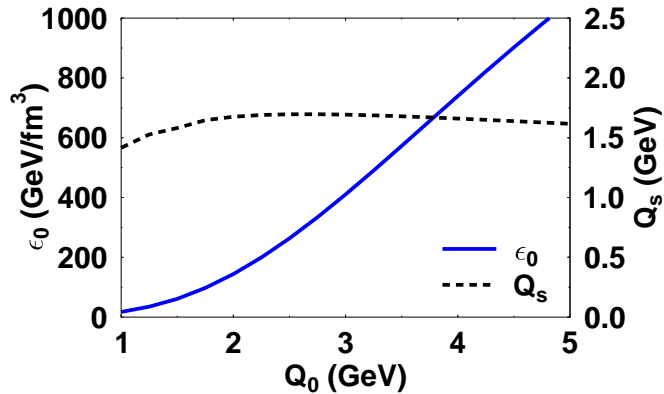,width=9cm}
\caption{Initial energy density $\epsilon_0$ of the gluon field at $\tau\to 0$
and saturation scale $Q_s = 1/R_c$ as a function of the cutoff $Q_0$.  
}
\end{center}
\end{figure}

Fig.\ 1 shows the dependence of the results on the (unphysical) cutoff
$Q_0$. The (physical) saturation scale $Q_s = R_c^{-1}$ is independent of 
$Q_0$ as it should be, while the initial energy density $\varepsilon$ 
($\epsilon_0$ in the plot) grows with $Q_0$. 
For a rather reasonable value $Q_0 = 2.5$ GeV the estimated initial energy 
density is roughly $\varepsilon = 260$ GeV/fm$^3$.

\ack
The author would like to thank his collaborators, J I Kapusta and Y Li,
and the organizers of QM 2006 for a memorable conference. This
work was supported in part by DOE grants DE-FG02-87ER40328,
DE-AC02-98CH10886, RIKEN/BNL and the Texas A\&M College of Science.

\end{document}